\documentclass[useAMS,usenatbib,usegraphicx]{mn2e}
\usepackage{multirow}
\title[The peculiar ringed galaxy: PGC 1000714]{A photometric study of the peculiar and potentially double ringed, nonbarred galaxy: PGC 1000714}

\author[B. Mutlu Pakdil et al.]{
{Burcin Mutlu Pakdil$^{1,2}$\thanks{E-mail: mutlu007@umn.edu}, Mithila Mangedarage$^{1}$, Marc S. Seigar$^{1}$, and Patrick Treuthardt$^{3}$}\\
$^{1}$Department of Physics \& Astronomy, University of Minnesota Duluth, 1023 University Drive, Duluth, MN 55812-3009, USA\\ 
$^{2}$Minnesota Institute for Astrophysics, University of Minnesota Twin Cities, 106 Pleasant St. SE, Minneapolis, MN 55455, USA\\
$^{3}$Astronomy \& Astrophysics Research Laboratory, North Carolina Museum of Natural Sciences, 11 W.\ Jones Street, Raleigh,\\
NC 27601, USA}

\begin{document}
\date{}
\pagerange{\pageref{firstpage}--\pageref{lastpage}} \pubyear{2014}

\maketitle

\label{firstpage}
\begin{abstract}
We present a photometric study of PGC 1000714, a galaxy resembling Hoag's Object with a complete detached 
outer ring, that has not yet been described in the literature. Since the Hoag-type galaxies are extremely 
rare and peculiar systems, it is necessary to increase the sample of known objects by performing 
the detailed studies on the possible candidates to derive conclusions about their nature, evolution, and systematic properties.
We therefore performed surface photometry of the central body by using the archival near-UV, infrared data and 
the new optical data (\textit{BVRI}). This current work has revealed for the first time an elliptical galaxy with two fairly round rings. 
The central body follows well a r$^{1/4}$ light profile, with no sign of a bar or stellar disc. By reconstructing the observed 
spectral energy distribution, we recover the stellar population properties of the central body and the outer ring. Our work 
suggests different formation histories for the galaxy components. Possible origins of the galaxy are discussed, and we conclude that 
a recent accretion event is the most plausible scenario that accounts for the observational characteristic of PGC 1000714.
\end{abstract}


\section{Introduction}

There are different types of galaxy rings such as polar \citep[e.g., A 0136-0801,][]{Whitmore1990}, collisional (e.g., Cartwheel Galaxy), 
accretion \citep[e.g., IC 2006,][]{Schweizer1989}, and the more familiar rings formed due to secular evolution \citep[e.g., NGC 1291,][]{ButaCombes1996}. 
These latter rings are commonly found in barred disc galaxies, though they are seen in non-barred disc galaxies as well \citep{Grouchy2010}. 
The origin of rings are generally related to either slow secular evolution or environmental processes. \citet{ButaCombes1996} pointed out 
the possibility that non-barred ringed galaxies may include both internally and externally-generated rings. Hence, non-barred ringed galaxies 
are among the ideal galaxies to study the role of both the internal dynamics of galaxies and the physics of accretion/interaction mechanisms.
  
An especially interesting ring case is Hoag's Object \citep[PGC 054559;][]{Hoag1950} with its peculiar morphology: an elliptical-like core with 
a nearly perfect outer ring, and no signs of bar and stellar disc. Hoag-type galaxies, which bear strong resemblance to Hoag's Object, are 
extremely rare \citep{Schweizer1987} and their origin is still debated. \citet{Brosch1985} proposed the ring of Hoag's Object 
was formed in a similar manner to rings in barred galaxies (i.e., the bar instability scenario involving slow internal motions). 
However, \citet{Schweizer1987} showed that the inner core is a true spheroid, not a disc, and suggested a major accretion event where a spheroid galaxy 
accreted and gas transferred from a colliding and/or a small companion galaxy into a ring. They attributed the absence of a merging signature to aging 
(2-3 Gyrs) of the accretion event. Both \citet{Brosch1985} and \citet{Schweizer1987} ruled out the possibility of Hoag's Object being a classical 
collisional ring galaxy seen face-on. As \citet{Brosch1985} pointed out, this scenario requires a high-level tuning of parameters such that a companion 
would have to pass through very close to the center of the disc and almost perpendicular to the plane with its relative velocity closely aligned into 
the line of sight. If all of these conditions were not met, the resulting galaxy would be a non-uniform or elliptical ring with an off-centered central 
object, so he narrowed down the probability of forming Hoag's Object via this mechanism to $10^{-5}$. Furthermore, \citet{Schweizer1987} discussed, had Hoag's 
Object been formed by this mechanism, it would have implied a relative velocity of order 100\,km\,s$^{-1}$ of the two galaxies and yet the relative 
velocity between the core and the ring is almost zero. \citet{Whitmore1990} considered this galaxy as a system related to Polar Ring Galaxies (PRGs), emphasizing
that Hoag-type galaxies are unlikely PRGs, but they may have similar evolutionary histories since they share some characteristics with them.
A recent study by \citet{Finkelman2011} considered accretion of gas from the intergalactic medium (IGM) on a pre-existing elliptical galaxy to explain both 
the peculiar structure and the kinematics of Hoag’s Object.

\citet{Wakamatsu1990} examined a list of Hoag-type galaxies and showed that Hoag-type galaxies often show elongated cores with disc-like features 
(e.g., NGC 6028), and probably represent an evolved phase of barred galaxies where the bar has been almost completely destroyed. However, 
\citet{FinkelmanBrosch2011} focused on UGC 4599 as the nearest Hoag-type ring galaxy. Having failed to detect a bar or a central discy component, 
the authors rejected the hypothesis that UGC 4599 is a barred early-type galaxy where the bar dissolved over time, and suggested that the 
cold accretion of gas from the IGM can account for the observed peculiar properties of the galaxy, although \citet{Maccio2006} demonstrated that gas 
accretion on a spheroidal is more likely to produce a S0 galaxy, rather than an elliptical-like galaxy. While \citet{Moiseev2011} considered UGC 4599 
as a possible face-on PRG, \citet{FinkelmanBrosch2011} emphasized that face-on polar rings with a Hoag-like structure were rarely discovered, despite 
a dedicated search \citep[see][]{Taniguchi1986,Moiseev2011}, implying that ellipticals with round bulges and almost circular rings as Hoag’s Object are 
exceedingly rare. \citet{FinkelmanBrosch2011} also did not rule out the possibility that part of the gas was accreted from a close companion during 
the evolution of the galaxy. Qualitatively, all these described scenarios are capable of describing a number of peculiarities of Hoag-type galaxies; 
however, the small number of known objects do not provide definite conclusions about their nature, evolution, and systematic properties. 
Therefore, it is necessary to increase the sample of known objects of this type by performing the detailed studies on the possible candidates. 

In this paper, we analyse images of PGC 1000714 (2MASX J11231643$-$0840067), which appears to be a genuine ring galaxy. While it was included 
in a number of catalogues, it has not yet described in the literature. Although it is classified as (R)SAa in NASA/IPAC Extragalactic Database (NED), 
this galaxy has a fair resemblance to Hoag’s Object in the optical and near-ultraviolet (NUV) bands, and that makes it a good target 
for a detailed study of Hoag-type galaxy structure. Such peculiar systems help our understanding of galaxy formation in general, since they represent 
extreme cases, providing clues on formation mechanisms. Since the photometric properties retain enough memory of the evolutionary processes that 
shaped the galaxies, one can retrieve valuable information about the nature of the structural components via a photometric study. Therefore, we perform 
surface photometry of the galaxy using both archival data (in NUV and infrared) and new data acquired in the optical band (\textit{BVRI}). 
By reconstructing the observed spectral energy distribution (SED) for the central body and the ring, we recover the stellar population properties of 
the galaxy components. 

This paper is organized as follows. Section 2 describes the general properties of the galaxy. After giving a description of the imaging data in 
Section 3, we present the methodology and detail results of isophotal fitting in Section 4. To further enhance the visibility of the outer ring, we perform 
a two-dimentional (2D) image analysis and present the results of the photometric decomposition in Section 5. After 
focusing on the rings in Section 6, we examine the stellar population properties in Section 7. Then, we discuss the implications of the observational 
evidence in Section 8 and finally summarize our main results in the conclusion.

Throughout the paper, unless specified otherwise, we use a standard cosmology of $H_{0}=73$ \,km\,s$^{-1}$\,Mpc$^{-1}$, $\Omega_{\Lambda}=0.73$ and 
$\Omega_{M}=0.27$ and magnitudes in the Vega system. 

\section{Observational Data}

The NUV imaging analysed here is obtained from archival GR6/GR7 data release of the GALEX All-Sky 
Imaging Survey. We use the background subtracted intensity map image corrected for the relative response. The image covers a 
field of 2.5 arcmin radius on the sky with a scale of 1.5 arcsec pixel$^{-1}$. The seeing full width at half-maximum (FWHM) is 
5.3 arcsec. Due the large seeing FWHM, the GALEX image is not very suitable for a detailed study of the inner part of the galaxy. 
Therefore, we use these data to investigate the outer ring. 
 
The optical images were observed at the 2.5-m du Pont telescope at the Las Campanas Observatory on 19 January 2006 with the Direct CCD 
Camera during dark time. The Direct CCD Camera is a 2048x2048 camera, with a pixel scale of 0.259 arcsec pixel$^{-1}$ resulting in a 
530 arcsec x 530 arcsec field of view. The image of PGC 1000714 just happened to appear in the background 
of another galaxy being observed, off to the bottom right of the field of the camera. This resulted in part of the galaxy in question 
falling off the side of the chip. Total exposure times were 2$\times$360 s in the \textit{B} band, 2$\times$180 s at \textit{V}, 2$\times$120 s 
at \textit{R}, and 2$\times$180 s at \textit{I}. The seeing measured in this particular set of images were 1.17, 1.11, 
1.01, and 0.96 arcsec in the \textit{B}, \textit{V}, \textit{R}, and \textit{I} bands, respectively. 

For each filter, 20 bias frames were taken and combined using the ZEROCOMBINE task in IRAF.  
The ccdproc command was then used to remove the bias level. The images were flatfielded using twilight flats. Several flat-field exposures 
were taken and combined using the FLATCOMBINE task, and ccdproc was then run to flatten the science images. Finally, the images were 
combined using the IMCOMBINE task by taking an average of the images and rejecting cosmic rays.  A mask file was created that contained 
positions of bad pixels and columns and the IMREPLACE task was used to account for these.  Finally, the imtranspose command was used 
to correct the orientation of the chip, so that North is up.

The images were taken during photometric conditions. Standard stars from the Landolt Equatorial Standard Stars Catalog \citep{Landolt1992} were observed 
at different airmasses through the night in all four filters.  We found that an aperture size of about 4$\times$ the image full width half maximum (FWHM) 
worked best for performing aperture photometry of the standard star images.  The photometric calibration was then performed using the DAOPHOT 
package within IRAF. Due to the high resolution and the small seeing FWHM, our work mostly relies on the optical data.

To determine the sky level, we apply fitting to the original image in linear steps of 10 pixels ($\sim2.6$ arcsec)
between successive ellipses. The program is forced to extend the fit well beyond the galaxy with the fixed ellipticity ($\epsilon$) 
and position angle (PA) of the outer ring. By checking the counts at these ellipses as a function of radius, we find that the ellipse 
at 40 arcsec leaves the galaxy and enters the background region, which is constant flux. After visual confirmation that these ellipses 
are really outside the galaxy, the mean and standard deviation of the fluxes between 42 and 78 arcsec are used as the final 
sky values (see Fig~\ref{fig1}). This method is very similar to the one presented by \citet{PohlenTrujillo2006}. 
As the surface photometry is sensitive to the adopted sky value, we check the sky values in a few ways. First, we calculate the mean 
of $\sim5$ median fluxes that were obtained from 10 pixel x 10 pixel boxes we positioned near the galaxy-free corners of each image. 
Second, we use the IRAF DAOEDIT task to estimate sky and sky-sigma of several stars. Finally, we check the most common pixel 
value by using the mode in IMSTAT task. Results from these different methods agree with the adopted values within the 2$\sigma$ 
uncertainty, which means the sky values and the uncertainties we adopted are well defined. 

\begin{figure}
  \caption{This figure illustrates our sky estimation. The isophotal intensity in counts of fixed ellipses fitted every 10 pixels. The region between the vertical dashed 
lines is used to estimate the sky value, which is indicated as horizontal dashed line together with the two dotted lines at $\pm1\sigma$.}
  \label{fig1}
\end{figure}

The near-infrared (NIR) images come from the Two Micron All Sky Survey (2MASS), which covers the entire sky in the 
filters \textit{J} (1.25\micron), \textit{H} (1.65\micron) and \textit{Ks} (2.17\micron). We use the image of a square field 
of view of $4\times4$ arcmin$^{2}$ on the sky with a scale of 1 arcsec pixel$^{-1}$. The seeing FWHM of the \textit{J, H, Ks}-band 
images is 2.48, 2.39, and 2.54 arcsec, respectively. Our calculations rely on the photometric zero point given in 
the keyword MAGZP in the header of each image, specifically 20.81, 20.41, and 19.92 mag. The sky values are simply taken from 
the keyword SKYVAL in the FITS header of each image. Unfortunately, the 2MASS survey has a weak sensitivity to low surface 
brightness due to the high brightness of the night sky in the NIR range and short exposures. For this reason, the part 
of the galaxy beyond the half light radius are unseen beyond the isophotes fainter than $\umu_{K}=$20 mag arcsec$^{-2}$. 

We also use the mid-infrared (MIR) images from the \textit{Wide-Field Infrared Survey Explorer (WISE)}, which covers the sky in four bands
centered at 3.4\micron (W1), 4.6\micron (W2), 12\micron (W3), and 22\micron (W4) with an angular resolution of 6.1,
6.4, 6.5, and 12.0 arcsec, respectively. We utilize the intensity images produced by match-filtering and coadding multiple 7.7 s (W1 and W2) 
and 8.8 s (W3 and W4) single-exposure images. The pixel scale of the images is 1.375 arcsec pixel$^{-1}$. The sky values are estimated with the same method 
used in the optical band. 

\section{General Properties}

The images of PGC 1000714 in different filters are shown in Fig~\ref{fig2}. The appearance of the galaxy 
in the NUV and optical band resembles to an almost round elliptical object with a detached outer ring, fairly similar to Hoag's Object. 
The outer ring is more clearly visible in the NUV and \textit{B}-band images. While still visible in the \textit{I} band, 
it disappears in the \textit{J} band. Both the infrared and optical images clearly show that the central body is the dominant luminous component
but the central body disappears in the W4 image.

 \begin{figure*}
  \caption{The images of PGC 1000714 in the NUV,\textit{B, I, J} and W3-band are shown from left to right, respectively.
For each image, the size is 75 arcsec x 90 arcsec, and the north is up while the east is on the left.
The appearance of the galaxy in the NUV and optical bands is fairly similar to Hoag's Object.}
  \label{fig2}
\end{figure*}

We investigate the galaxy by deriving the high-frequency residual image, which is produced by taking the ratio of the original image to 
its median filtered image, which is computed with the FMEDIAN package in IRAF, where each original pixel value is replaced with 
the median value in a rectangular window. Using a small window size highlights the prominent galaxy substructure while using a larger window
size highlights the prominent large-scale galaxy components. First, we use a window size of 5 x 5 pixels in the optical bands, and have failed
to detect any prominent substructure. Then, we use a window size of 301 x 301 pixels in the \textit{B} band, and highlight the central 
body and the outer ring (see Figure~\ref{fig3}). Their extensions are roughly estimated by 'eye'. The outer ring has an orientation and ellipticity similar to the core, 
with slightly higher values: $PA_{ring}=60\degr$, $PA_{core}=50\degr$, $\epsilon_{ring}=0.20$, $\epsilon_{core}=0.15$. Note that, in Fig~\ref{fig1}, 
the central body extends through the outer ring which appears as a bump starting at 20 arcsec and fades away at 40 arcsec. 
 
\begin{figure*}
  \caption{Left-hand panel shows the high-frequency residual \textit{B}-band image of PGC 1000714, which is obtained using a window size of 301 x 301 pixels 
in order to get the optimum enhancement of the galaxy prominent structures. Any prominent component appears as a region with a value greater than 1. The bright 
central object appears as star-like (or bulge-like). The length of the semimajor axis (SMA) for the yellow ellipse is 18 arcsec, while those of the white ones 
are 25 and 41 arcsec. All three ellipses share the same center. Right-hand panel shows the residual image that was produced by fitting ellipses to 
the core and subtracting the BMODEL image from \textit{B}-band image. Two background sources can be seen in the region where the bulge was subtracted, 
within the outer ring. The core does not show any underlying structure.}
  \label{fig3}
\end{figure*}

The general information reported by NED is listed in Table 1. The redshift of PGC 1000714 is $0.0257\pm0.0001$ ($7717\pm38$\,km\,s$^{-1}$; NED).
We check the number of neighbours of PGC 1000714 by searching NED for nearby galaxies with known and similar redshifts. 
A similar search was done by \citet{FinkelmanBrosch2011} to identify the environment of the galaxy in question. We search 
within a radius of 120\arcmin of the object and a redshift between 6700 and 8700\,km\,s$^{-1}$. We find that PGC 1000714 does 
not lie within the boundary of a recognized cluster of galaxies and its nearest neighbour GALEXASC J112227.40-081817.0 is 
about 24.95 arcmin away. Therefore, we characterize its neighborhood as a low-density environment.

\begin{table}
\centering\large
\caption{General information for PGC 1000714 from NED}
\label{T1}
\begin{tabular}{l|c} 
          \hline
          {}                 &  PGC 1000714                 \\
          \hline
          RA(J2000)       & $11^{\rm h}23^{\rm m}16.44^{\rm s}$   \\
          Dec.(J2000)     & $-08^{\rm d}40^{\rm m}06.5^{\rm s}$   \\          
          Radial Velocity (km\,s$^{-1}$) & $7717\pm38$         \\ 
          z               & 0.025741                         \\
          Distance (Mpc)  &  110                             \\
        \hline
        \end{tabular}
\end{table} 

\section{Isophotal Analysis}
\subsection{Ellipse Fitting}

We analyse the isophotal structure and luminosity profile of the galaxy using the IRAF task ELLIPSE, which follows the 
procedure outlined by \citet{Jedrzejewski1987}, that fits a set of elliptical isophotes to an image. Before the fitting, all 
foreground stars and other contaminants have been identified and masked by using SEXTRACTOR \citep{BertinArnouts1996}. 
$\epsilon$ and PA are allowed to vary. To look for a possible varying center, the center is also left as a free parameter 
during ellipse fitting. However, the typical shifts obtained were less than a pixel ($<0.26$ arcsec). In order to search for hidden 
structures in the central core, we restrict our fitting to the radius of $r=18$ arcsec to avoid light contamination
by the outer ring (see Figure~\ref{fig3}, the yellow ellipse). The BMODEL task is then used to construct the core model. Subtraction 
of the model revealed that there are two background/foreground sources located between the radius of $\sim5$ and 
$\sim10$ arcsec: one is in the north-west, and other is in the south of the centre. We mask these source manually, and repeat 
the ellipse fitting. This time, subtraction of the model from the actual image results in a very clean image (see the right-panel in Fig~\ref{fig3}). 
The residual image shows no sign of an underlying substructure in the core. Although the core is modelled quite well, there is some asymmetry 
that is only visible on the scale of the point spread function (PSF), and this may be completely caused by the seeing effect. 

The geometric properties derived from the optical, W1 and W2 images are shown in Fig~\ref{fig4}. The results from the optical 
data are remarkably in good agreement. The $\epsilon$ and PA profiles display a clear systematic behaviour and follow a well-defined pattern. 
The central body is nearly round with an $\epsilon$ value of $\sim0.1\pm0.05$ with the major axis at PA$=50\pm10$ \degr.
The isophotal shape parameters \citep[\textit{c$_{4}$} and \textit{s$_{4}$}; see e.g.,][]{Milvang-JensenJorgensen1999} stay very close to zero 
for the inner roughly 5 arcsec, as expected for an elliptical or round object. The \textit{B}-band \textit{c$_{4}$} profile does not experience 
any significant variation, implying the core is featureless at shorter wavelengths. At longer wavelengths, the \textit{c$_{4}$} 
values become slightly higher, that might suggest an old underlying substructure. Note that the \textit{c$_{4}$} is an indicator of a disky 
($c_{4} > 0$) or a boxy ($c_{4} < 0$) structure \citep{Milvang-JensenJorgensen1999}. However, the absence of a systematic behaviour 
in the optical \textit{c$_{4}$} profiles (such as becoming negative or positive) is consistent with the absence of a bar-like feature.
Although the \textit{c$_{4}$} profile becomes positive in the W1 and W2, this can be the result of the seeing effect 
instead of a hidden structure. The seeing FWHM in our \textit{MIR} data is much larger than the one in our optical data. The seeing disc 
has severe influence within the radius of $2\times$FWHM \citep{Graham2001a,Huang2013}. For the MIR data, the $2\times$FWHM 
roughly corresponds to 12 arcsec which covers most of data points. Therefore, we cannot draw any firm conclusion from the MIR data. 
The \textit{BVRI}-band images have the largest ratio ($R_{e}/FWHM$; see Table 2 for $R_{e}$), hence the derived properties are more reliable 
when compared to those derived from the MIR images. In any case, the profiles are not very reliable near the centre. This is because 
the inner light gets distributed to larger radii due to the PSF. This can be the reason of the peak in the $\epsilon$ which is observed just 
outside the FWHM of the PSF, in the \textit{BVRI}-bands. Similar to the \textit{c$_{4}$}, the \textit{s$_{4}$} values within the radius of 
5 arcsec are all consistent with zero value. Outside this inner region, the \textit{s$_{4}$} values get as high as that of the \textit{c$_{4}$}. 
These higher \textit{s$_{4}$} values might be caused by a spiral or a ring \citep[e.g.,][]{Gadotti2007}. 

Based on the optical data, we are inclined to believe that the central body is truly not barred. We investigate the core further with 2D image decompositions to 
check if the relatively higher values of the \textit{c$_{4}$} and the \textit{s$_{4}$} at the longer wavelengths are caused by an underlying substructure (see Section 5). 

\begin{figure*}
\centering
  \caption{Left-top panel shows the variation of ellipticity. We only display the data whose error is less 
than 0.04 in $\epsilon$. Left-bottom panel shows the variation of position angle (PA). We only display the data whose error 
is less than 20\degr in PA. Both profiles are fairly smooth. Right-top panel shows the variation of \textit{c$_{4}$}.
We only display the data whose error is less than 0.02 in \textit{c$_{4}$}. \textbf{Right-Bottom} panel shows the variation of 
\textit{s$_{4}$}. We only display the data whose error is less than 0.02 in \textit{s$_{4}$}. Both \textit{c$_{4}$} and \textit{s$_{4}$} 
profiles are close to zero within inner 5 arcsec, then they start to experience some higher variations toward the outer radii. 
The FHWM of the PSF for each band is shown with the red vertical dotted line.}
  \label{fig4}
\end{figure*}

\subsection{Radial Profile}
The colour profiles are derived from the ellipse fits in each band separately. The NUV profile is noisy compared to other profiles. Both the NUV 
and infrared colours suffer significantly from the large seeing effect within $2\times$FWHM. Only the optical colours are well estimated. A careful inspection of 
the optical profiles reveals a fairly constant colour index (see Fig~\ref{fig5}). We find that the optical colour index at the inner region ($\sim1$-$12$ arcsec) 
is fairly constant, but starts to vary slightly outside $\sim12$ arcsec.

\begin{figure*}
  \caption{The colour gradients are shown: the NUV-$B$, optical ($B$-$V$, $B$-$R$, $B$-$I$),
NIR ($I$-$J$, $I$-$H$, and $I$-$Ks$), and MIR ($I$-W1, $I$-W2, and $I$-W3) colour plots. In the \textit{JHKs} bands, the outer parts of the galaxy 
are unseen beyond $\sim8$ arcsec. The vertical red dashed line indicates the FWHM of the PSF in each band, e.g. top-left: $\sim5$ arcsec for NUV, $\sim1$ arcsec 
for optical, $\sim2.5$ arcsec for NIR, and $\sim6$ arcsec for MIR. The optical colour indices of the galaxy present fairly constant radial profile, 
the others are not reliable due to the larger seeing FWHM.}
  \label{fig5}
\end{figure*}

To examine the light distribution along the core, the core profiles in the \textit{BVRI} bands are fitted to Sersic r$^{1/n}$ profile \citep{Sersic1968} 
with the least-squares minimization IDL package MPFIT. The optical images are used due to the smallest seeing effect. 
The Sersic model converges toward the de Vaucouleurs law, yielding a value for n that is quite close to 4. To look for a possible disc
component, we also try to fit to the Sersic bulge $+$ exponential disc profile, however this model cannot fit the data,
indicating the absence of a stellar disc structure. The best-fit in the \textit{B} band indicates that the single r$^{1/4}$ component matches well the light profile 
over a wide $\sim6$ mag range. Fig~\ref{fig6} shows the surface brightness profile of the core with Sersic profile fit in each optical band as a function of radius.
The lower panel shows the residuals from the best-fitting model profile to highlight deviations from the model. While the \textit{B}-band residuals are consistent 
with the zero value, the residuals toward the longer wavebands hint about the presence of a reddish substructure. We investigate these residuals further with 2D image 
decompositions by using the derived structural parameters from the 1D fitting (see Table 2 and Section 5).    

\begin{figure*}
  \caption{The best Sersic model fit for the \textit{BVRI}-bands. The best fit r$^{1/n}$ model in the \textit{B} band gives $n=4$, and the best 
fits in other bands give $n$ values very close to 4. $R_{e}$ is the effective half light radius of the galaxy. The lower panels show the residuals from 
the best-fit Sersic profile to highlight deviations from the model.}
  \label{fig6}
\end{figure*}

\section{2D Image Analysis}
After extracting the 1D profiles, we further investigate the galaxy by deriving a 2D $B$-$I$ colour index map (see Fig~\ref{fig7}). To correct for the PSF differences in each band, 
we use IRAF task GAUSS to convolve the \textit{B}-band image with the PSF in the \textit{I}-band, and convolve the \textit{I}-band image with the PSF in the \textit{B} band. 
Then, the colour map is obtained by transforming the images from the intensity units into magnitude units, and correcting for the corresponding zero-point 
magnitudes, and then simply subtracting these two images. The 2D map reveals the existence of a diffuse inner ring with no sign of a bar. 
The existence of the inner ring  explains the characteristic residuals seen in Fig~\ref{fig6}. In addition, it is consistent with the \textit{c$_{4}$} and the \textit{s$_{4}$} 
profiles that we derived in Section 4.1. 

\begin{figure}
  \caption{The 2D $B-I$ colour index map is shown. Note that the light green colour scale reveals the existence of a diffuse inner ring.
 Both the inner and outer ring are pointed by the white arrows. The size is 79\arcsec x 110\arcsec, and the north is up while the east is on the left.}
  \label{fig7}
\end{figure}

The 2D image decomposition is performed by using GALFIT profile fitting code \citep{Peng2010}. The outer and inner ring are 
carefully masked to prevent an overestimation of the outer parts of the galaxy. Then, the galaxy is modelled by using the same two 
functions, while taking the PSF into account. Since GALFIT requires initial guess parameters to fit the data, we exploit the 
results from the 1D fitting to provide suitable initial guesses. We first allow them to vary, and then fix them to be able to compare 
the residuals. Similar to our previous fitting, the best fit of Sersic function yields a value for $n$ that is quite close to 4 when the input parameters 
are allowed to vary. The Sersic bulge $+$ exponential disc function does not fit the data. The residual images generated by GALFIT 
are shown for in Fig~\ref{fig8} for the optical bands. Table 2 lists our fitting results from GALFIT for two cases (fixed 
and free parameters), which are quite consistent with each other. The derived structural parameters are in very good agreement in all bands.

\begin{figure*}
  \caption{The 2D image decomposition is shown for the \textit{BVRI} bands (from left to right respectively). For each image, the size is 66\arcsec x 117\arcsec, 
and the north is up while the east is on the left. Top panel displays images of the original data, middle panel displays the residuals from Sersic bulge model 
generated by {\tt GALFIT}, in which the initial parameters are allowed to vary. Bottom panel displays the residuals from the Sersic bulge model, but this time 
the initial parameters are fixed to the 1D fitting results obtain from {\tt MPFIT}. While the presence of the inner ring is almost undetectable in the \textit{B}-band residuals,
it gradually becomes prominent in the \textit{VRI} band residuals.}
   \label{fig8}
\end{figure*}

\begin{table*}
\begin{minipage}{160mm}
\centering\small
\caption{Structural parameters obtained by GALFIT}
\label{T2}
\begin{tabular}{lccccc} 
          \hline
          {} & \textit{NUV}   & \textit{B} &  \textit{V}  &  \textit{R} &  \textit{I}\\
          \hline
          The initial parameters are fixed to the 1D fitting results. \\
          \hline
          CORE\\
          $m_{total}$ (mag)               & $17.82\pm0.22$   & $16.39\pm0.01$   & $15.46\pm0.01$ & $15.25\pm0.02$ & $14.12\pm0.02$ \\
          $\mu_{e}$ (mag arcsec$^{-2}$)   & $24.47\pm1.02$   & $23.07\pm0.03$   & $22.10\pm0.04$ & $21.81\pm0.08$ & $21.14\pm0.09$ \\
          $R_{e}$ (arcsec)               & $6.63\pm3.76$    & $4.75\pm0.08$    & $4.68\pm0.09$  & $4.50\pm0.17$  & $5.46\pm0.25$ \\
          $n$                           & 4.00\footnote{The core is fitted by the r$^{1/4}$ model.} & $4.00\pm0.01$    & $4.32\pm0.07$  & $4.52\pm0.14$  & $5.02\pm0.16$ \\
          $b/a$                         & $0.51\pm0.14$   & $0.91\pm0.01$    & $0.89\pm0.01$  & $0.86\pm0.01$  & $0.85\pm0.01$ \\
          P.A. (degree)                 & $23.73\pm0.06$  & $48.38\pm1.30$   & $48.33\pm0.99$ & $48.44\pm0.64$ & $48.91\pm0.62$ \\
 %         $\chi^{2}$  & ... & 1.117 & 1.130 & 1.127 & 1.127 \\
          INNER RING\\
          $m_{total_{E}}$ (mag)           & ...            & $19.95\pm0.13$ & $19.06\pm0.41$ & $18.55\pm0.12$ & $17.36\pm0.17$\\
          $m_{total_{C}}$ (mag)           & ...            & $20.00\pm0.16$ & $19.07\pm0.39$ & $18.60\pm0.14$ & $17.37\pm0.20$\\
          OUTER RING\\
          $m_{total_{E}}$ (mag)          & $17.39\pm0.07$ & $18.05\pm0.13$ & $17.54\pm0.09$ & $17.89\pm0.25$ &  $17.08\pm0.61$ \\
          $m_{total_{C}}$ (mag) &        $17.35\pm0.06$ & $18.04\pm0.15$ & $17.51\pm0.10$ & $17.89\pm0.30$ &  $17.06\pm0.63$ \\
          \hline
          The initial parameters are allowed to vary.\\
          \hline
          CORE\\
          $m_{total}$ (mag)              & $17.65\pm0.10$  & $16.33\pm0.05$     & $15.32\pm0.05$        & $15.17\pm0.08$          & $14.00\pm0.10$ \\
          $\mu_{e}$ (mag arcsec$^{-2}$)  & $24.82\pm0.62$  & $23.27\pm0.15$   & $22.60\pm0.16$        & $22.11\pm0.25$          & $21.53\pm0.35$ \\
          $R_{e}$ (arcsec)              & $8.03\pm2.28$   & $5.43\pm0.41$    & $6.20\pm0.50$         & $5.41\pm0.69$           & $6.96\pm1.26$ \\
          $n$                          & $3.57\pm1.31$   & $4.02\pm0.24$    & $4.85\pm0.26$         & $4.65\pm0.44$           & $5.22\pm0.59$ \\
          $b/a$                        & $0.50\pm0.08$   & $0.88\pm0.01$      & $0.86\pm0.01$         & $0.85\pm0.01$           & $0.84\pm0.01$ \\
          P.A. (degree)                & $25\pm6.37$     & $48.68\pm1.01$   & $48.24\pm0.79$        & $48.22\pm0.65$          & $48.98\pm0.60$\\
%          $\chi^{2}$ & 1.003      & 1.114           & 1.129        & 1.122 & 1.122\\
          INNER RING\\
          $m_{total_{E}}$ (mag)          & ...            & $20.85\pm0.22$ & $20.03\pm0.20$ & $19.32\pm0.28$ & $18.30\pm0.35$\\
          $m_{total_{C}}$ (mag)          & ...            & $20.85\pm0.25$ & $20.22\pm0.28$ & $19.38\pm0.30$ & $18.32\pm0.34$\\
          OUTER RING\\
          $m_{total_{E}}$ (mag) & $17.41\pm0.03$  & $18.14\pm0.06$ & $17.91\pm0.03$ & $18.00\pm0.17$ & $17.46\pm0.20$ \\
          $m_{total_{C}}$ (mag) & $17.40\pm0.03$  & $18.12\pm0.08$ & $17.91\pm0.06$ & $18.01\pm0.22$ & $17.50\pm0.26$ \\         
          \hline
        \end{tabular}
\end{minipage}
\end{table*} 

\section{Rings}

The outer ring can be seen well in the NUV and \textit{BVRI} bands (see Fig~\ref{fig2}). While NIR imaging is not optimal for studying 
the outermost part of the galaxy, the outer ring cannot be detected in the MIR data. In Section 4.1, we aim to study the geometric properties 
of the core (e.g. the $\epsilon$ and PA profiles), so we allow $\epsilon$ and PA to vary. Due to the clumpy and non-uniform structure of the ring,
the unconstrained ellipse fitting procedure cannot be applied on the outer ring. To study the structure of the outer ring, we repeat
the ellipse fitting, but this time we force the program to extend the fit well beyond the core with the fixed $\epsilon$ and PA of the outer ring, 
which is the ellipticity ($\epsilon=0.2$) and orientation (PA$=60$\degr) of the white ellipses in the Fig~\ref{fig3}. We find the mean surface brightness 
drops to $\umu_{B}=26.8$ mag arcsec$^{-2}$ at the faintest level of the core, at the radius of $\sim21.5$ arcsec. 
The outer ring has no sharp outer boundary and fades away to the radius of $\sim40$ arcsec. The peak surface
brightness of the outer ring is $\umu_{B}=25.9$ mag arcsec$^{-2}$ at $r=30$ arcsec and is fainter than that of Hoag's
object (\citet{Finkelman2011} reported the peak surface brightness of the outer ring for Hoag’s object is $\umu_{B}=24.6$ mag arcsec$^{-2}$ 
at $r=19.1$ arcsec). In the case of the NUV band, the core profile is fitted by the r$^{1/4}$ model. We did not fit the Sersic bulge model 
to the NUV data due to the large seeing effect. The light profile of the galaxy with the outer ring is shown in Fig~\ref{fig9}, where the lower panel 
shows the residual after the best-fitting core model in Fig~\ref{fig6} is subtracted. The residuals reveal an inner ring at radius between
$\sim8$ arcsec and $\sim15$ arcsec, which is very faint in the blue light, and an outer ring at radius between 20 and 40 arcsec, 
which is very bright in the NUV light. Unfortunately, we cannot verify the nature of the faint inner ring in our MIR 
data due to the large PSF effects.

\begin{figure*}
  \caption{The light profiles of the galaxy with the outer ring are shown for the NUV and \textit{BVRI} bands. 
The lower panels show the residual after the best-fitting core model in Fig~\ref{fig6} is subtracted. The residuals reveal an 
inner ring at radius between $\sim8$ and $\sim15$ arcsec: It is very faint in the blue light and 
gets brigher towards redder light. In addition, the residuals show an outer ring at radius between $\sim20$ 
and $\sim40$ arcsec: It is very bright in the NUV light and gets fainter towards redder light.} 
  \label{fig9}
\end{figure*}

To estimate the total intensity of the rings, we use the residual images generated by GALFIT (see Fig~\ref{fig8}). 
Using the residual images, total intensities of the outer ring are derived from the region between 20 and 
40 arcsec while the total intensity from the region inside 20 arcsec is adopted as that of the inner ring. The 
region is enclosed by the same white ellipses in Fig~\ref{fig3}. The total intensities of the rings are also derived 
from the region defined by circles. These intensities are then converted into magnitudes, using proper zero-point magnitudes. 
The derived magnitudes are listed in Table 2.    

\section{Stellar Populations}

SED fitting is a common technique for deriving galaxy properties, which relies on comparing 
the observed SED to a set of model SEDs and searching for the best match. Since the physical properties of the model 
galaxies are known, this knowledge can be used to recover the properties of an observed galaxy. In order to study the 
stellar populations of the core and the outer ring, we use HYPERZ SED fitting code, version 1.2 \citep{Bolzonella2000}. 
Due to the large uncertainty on the nature of the inner ring in the infrared light, we did not attempt to constraint its age. 
The observed SED for the bulge and the outer ring is defined by the mean of the magnitudes listed in Table 2, and the standard 
deviation of the values is used to estimate the uncertainty (see Table 3).  

\begin{table}
\begin{minipage}{80mm}
\centering
\caption{The observed SED of the galaxy components in units of magnitude.}
\label{T3}
\begin{tabular}{lccc} 
          \hline
          {}                                 &Filters   &Bulge             &Outer ring\\
          \hline
          Near-UV                           & NUV  &$17.74\pm0.24$    &$17.39\pm0.11$\\
          \hline
          \multirow{4}{*}{Optical}          &  \textit{B}    &$16.36\pm0.06$    &$18.09\pm0.23$\\
                                            &  \textit{V}    &$15.39\pm0.11$    &$17.72\pm0.27$\\
                                            &  \textit{R}    &$15.21\pm0.10$    &$17.95\pm0.48$\\
                                            &  \textit{I}    &$14.06\pm0.13$    &$17.28\pm0.97$\\
          \hline
          \multirow{3}{*}{Near-IR}          &  \textit{J}    &$13.34\pm0.52$    &...\\
                                            &  \textit{H}    &$12.80\pm0.44$    &...\\
                                            &  \textit{Ks}   &$12.51\pm0.34$    &...\\
          \hline
          \multirow{4}{*}{Mid-IR\footnote{We make use of the photometric information from the WISE Source Catalog.}} & W1            &$12.56\pm0.35$    &...\\
                                            & W2            &$12.61\pm0.30$    &...\\
                                            & W3            &$11.64\pm0.44$    &...\\
                                            & W4            &...               &...\\
          \hline
        \end{tabular}
\end{minipage}
\end{table}

The SED fitting method is based on minimization of $\chi^{2}$, which is computed for templates covering a broad range 
of star formation modes, ages and absorption values. The default is an age grid of 51 values (minimum and maximum age are 
32 Myr and 19.5 Gyr, respectively). Galaxy age is constrained to be younger than the age of the Universe. 
The best-fitting solution yields the stellar age, star formation law, and dust reddening. Our set-up consists of the mean 
spectra of local galaxies from \citet{ColemanWuWeedman1980}, and \citet{BruzualCharlot1993} evolving sythesis models. 
We assume single burst, constant star formation and exponentially decaying star formation histories (SFH) where star 
formation rate (SFR) is defined as SFR$(t)=(1/\tau)\exp(-t/\tau)$ in which the $\tau$ parameter is the 'time-scale' 
when the star formation was most intense with $\tau=1,2,3,5,15,30$ Gyr, as included in {\tt HYPERZ} package.
We use the attenuation law from \citet{Calzetti2000}, and assume solar metallicity and a \citet{Salpeter1955} initial 
mass function (IMF) with lower and upper mass cut-offs of 0.1M$_{\sun}$ and 100M$_{\sun}$. In addition to this set-up, 
we also use a self-consistent set of templates, where the evolution in metallicity of stellar population is explicitly 
taken into account \citep[see][]{MobasherMazzei1999}, as included in HYPERZ package. The best-fitting solution for 
the bulge is obtained by an exponentially decaying SFH with $\tau=1$ Gyr with solar metallicity. The age of the best template 
for the bulge is 5.5 Gyr (falling between 4.5 Gyr and 6.5 Gyr in the age grid). The observed SED of the outer ring is well reproduced 
by a stellar population with solar metallicity by an instantaneous burst scenario. The age of the best template 
for the ring is 0.1278 Gyr (falling between 0.09048 Gyr and 0.18053 Gyr in the age grid). Fig~\ref{fig10} shows the observed SED of the bulge
and the ring with their best-fitting models. The identical results are obtained when different IMFs (namely \citet{MillerScalo1979} and 
\citet{Chabrier2003}) are used in the set-up. 

%\clearpage
\begin{figure*}
  \caption{The observed SED of the core and the ring (red circles) with their best-fitting models (black solid lines) are 
shown. The y-axes show the flux units in AB magnitudes.} 
  \label{fig10}
\end{figure*}

\section{Discussion}

We have shown earlier that PGC 1000714 presents an enhancement of the blue light in a fairly circular region centered on the core. 
The detailed photometry reveals an elliptical galaxy-like core with a diffuse reddish inner ring-shaped structure, but no sign of 
a bar or stellar disc. The SED fitting suggests that the galaxy is fairly young: the core being $\sim5.5$ Gyr old, and the detached 
outer ring being $\sim0.13$ Gyr old. The central body is well matched with an exponential decaying SFR with $\tau=1$ Gyr, that produces a 
reasonable fit of the photometric properties of elliptical galaxies in the local Universe. A shallow intrinsic metallicity gradient along the core, 
as implied by Fig~\ref{fig5}, is consistent with a major merger scenario, in which a merging event causes the turbulent mixing and flattens any 
possible ambient gradient present in the progenitor galaxies \citep{White1980}. It is unlikely that the core is the result of a monolithic gas collapse. 
According to this model, a rapid dissipative collapse forms stars while the gas sinks to the centre of the forming galaxy 
\citep{Larson1974,L1975,Carlberg1984,ArimotoYoshii1987}. The infalling gas which is chemically enriched by evolving stars contributes the metal-rich star 
formation in the centre, thus establishes very steep colour gradients \citep[and references therein]{Sanchez-Blazquez2006,Spolaor2009}. Our shallow colour gradient 
along the core contradicts the predictions of the monolithic collapse model. The SED fitting deduces clearly different ages, stellar characteristics and SFHs 
for the outer ring and the core. Based on the colour of the rings, we expect different formation histories for the inner and outer ring, however
we cannot determine the formation mechanism for the dilute inner ring with confidence based on the data presented in this paper. 
To say more about the formation of the inner ring, we need infrared data of PGC 1000714 with better resolution.

A number of ideas may be put forward to explain the outer ring in such a peculiar galaxy, but the observational results rule out 
most of these. In the widely accepted bar-driven ring model, one may expect that PGC 1000714 was a barred early-type galaxy once, 
where it experienced a strong bar instability in the disc, an outer ring formed in the process, and the bar and disc were 
subsequently destroyed. A vertical buckling instability in the disc may indeed weaken the bar and dissolve its outer half 
\citep{Martinez-Valpuesta2006} within a Gyr or less \citep{AthanassoulaMartinez-Valpuesta2008}. However, the buckling instability 
is not expected to lead to a complete bar dissolution \citep{Martinez-ValpuestaShlosman2004}. As reviewed by \citet{ButaCombes1996}, 
if a satellite merger is the driving force for the bar destruction, the effective dissolution of the bar may occur in about 20 Myr \citep{Pfenniger1991}. 
However, such a process is expected to cause some non-axisymmetric features on the central body, which we have not detected. 
If gas infall towards the center is the reason for the bar destruction, the decoupling of a secondary 
bar might occur in the dissolution of the primary one when there is a high accretion rate \citep{FriedliMartinet1993}. For a milder 
gas accretion rate, the time-scale for the bar dissolution can be very long due to self-regulation \citep{ButaCombes1996}. In addition, 
the bar destruction process affects the orbital structures and may cause lens formation \citep{Combes1996}, which is not detected in 
PGC 1000714. Moreover, such a process is also expected to create a light profile of the core with a low Sersic index of $n<2$ 
because pseudo-bulges are expected to retain fingerprints of their disky origin \citep{KormendyKennicutt2004}. Yet the core has 
no sign of a disc, and presents properties of a spheroid with a smooth r$^{1/4}$ light distribution for more than 5.0 mag. 
Note also that the rings of barred galaxies are much more luminous than that of PGC 1000714 (M$_{B}=-16.9\pm0.1$), 
e.g. M$_{B}\sim-20.5$, $-20.3$, and $-19.6$ for NGC 1291, NGC 2217, and NGC 2859. \citet{Debattista2004} used collionless 
N-body simulations to study the final properties of discs that suffer a bar instability, and failed to reproduce round pseudo-bulges 
inside low-inclination galaxies, as in PGC 1000714. As pointed out by \citet{Schweizer1987} for Hoag's Object, a disc is needed to 
give rise to a bar instability. However, it is worth mentioning that \citet{GadottideSouza2003} proposed a mechanism to form bars in 
spheroids through the dynamical effects of a sufficiently eccentric halo, without the need of a stellar disc. To say more about 
this model, the information about the halo structure of the galaxy is needed. 

The interactions of galaxies can be considered as an alternative mechanism for the outer ring formation. In the galaxy-galaxy collision 
model, PGC 1000714 can be considered as a classical ring galaxy with its companion superimposed. High-resolution numerical simulations 
of binary mergers of disc and early-type galaxies find most of the fast rotators formed due to major mergers have intermediate ellipticities 
between 0.4 and 0.6, and mostly contain a bar \citep{Bois2011}.  However, we find that PGC 1000714 has a fairly round central body with 
the fairly round inner ring with no trace of a bar. Moreover, the geometrical properties derived from the optical images stay constant along 
the central core (e.g., $\epsilon\approx0.1$ and PA$\approx50$ \degr). The smooth profile at the outer region imply that the face-on ring 
is in equilibrium with the host galaxy and that the system has settled to its current configuration. Since the initial conditions must be fine-tuned
in order to create a polar ring from a merger event \citep{Maccio2006}, it is highly unlikely this will create the peculiar properties of 
PGC 1000714 from a merger event. Therefore, we rule out the possibility of the galaxy having a history of a major collision. 

A close passage of a nearby galaxy or a tidally interacting companion can mimic the effect of a bar by creating a 
spiral structure or a ring in non-barred galaxies \citep{ElmegreenElmegreen1983,Combes1988,TutukovFedorova2006}. 
This non-axisymmetric presence can trigger a recent burst of star formation and create a UV-bright ring structure in the outer 
discs of spirals \citep{Thilker2007}. Although an orbiting small companion could create a detached outer ring, such an external 
perturbation could also destroy an existing ring, break a true ring into a pseudo-ring structure, or prevent it from occuring in 
the first place \citep{ButaCombes1996}. Together with our failure in finding a possible companion as well as detecting any tidal tail, 
shells or ripple signature brighter than $\umu_{B}\approx27$ mag arcsec$^{-2}$, it is unlikely that the tidal action is the origin of 
the outer ring in PGC 1000714. 

An accretion model can also explain the formation of a detached outer ring. This model is used to explain the formation of narrow PRGs, 
which share similar characteristics with PGC 1000714. As an example, the core of narrow PRG AM 2020-504 follows well the r$^{1/4}$ 
for a large range of magnitudes \citep{Arna1993,Iodice2002} and its polar ring appears as a peak on the underlying central body 
\citep{Iodice2014}. The age of the central body and the ring of AM 2020-504 (3-5 Gyr and 1 Gyr, respectively) are comparable to those 
of PGC 1000714 \citep{Iodice2014}. As a formation mechanism for  AM 2020-504, \citet{Iodice2014} suggested tidal accretion of material 
from outside by a pre-existing early-type galaxy since this process would not modify the global structure of the progenitor accreting 
galaxy, whose morphology remains spheroidal with same colours and age. Due to the similar characteristics, this argument is also 
valid for PGC 1000714. However, it is worth mentioning that the radius of the outer ring in PGC 1000714 is not small with respect to 
the semi-major axis radius of the central body as in the narrow PRGs, hence PGC 1000714 cannot be a narrow PRG with the polar structure 
moderately inclined to the line of sight (nearly face-on). The exponential-like light profile of the polar structure in the wide PRGs 
\citep{Reshetnikov1994,Iodice2002,Spavone2012} is also not consistent with that in PGC 1000714. On the other hand, 
the absolute magnitude of the outer ring (M$_{B}=-16.9\pm0.1$) is comparable to those of PRGs, e.g. M$_{B}=-18.2\pm0.5$, 
$-17.6\pm0.6$, and $-15.5\pm1.0$ for the wide PRG NGC 4650A, the wide PRG A0136-0801, and the narrow PRG ESO 415-G26, respectively 
\citep{Whitmore1987}. Peculiar galaxies with several decoupled ring-like structures are also observed, in which at least one of 
them is on the polar direction with respect to the central host, e.g. ESO 474-G26 \citep{Reshetnikov2005,Spavone2012}. While the central 
body of  ESO 474-G26 is an elliptical-like object (and almost round; $\epsilon\sim0.06$), it is much more brighter than that of PGC 1000714.
Also, the both rings in ESO 474-G26 have similar colour, and they are very irregular \citep{Spavone2012}. The simulations suggest 
that the structure of ESO 474-G26 could be a transient phase and not a stable dynamical configuration \citep{Reshetnikov2005,Spavone2012}.
\citet{Spavone2012} ruled out the gradual disruption of a dwarf satellite galaxy as a formation mechanism for this multiple ring structure. 
However, note that the rings of PGC 1000714 which have different colours require different formation mechanisms, therefore the 
argument made by \citet{Spavone2012} is not valid for it. Also, unlike ESO 474-G26, the smooth geometric profile at the outer region of 
the core imply that the face-on ring is in equilibrium with the host galaxy and that the system has settled to its current configuration. 
No  kinematic data is available for PGC 1000714, therefore we cannot verify whether it is a PRG or not. 
However, it shares several characteristics with them, therefore they may be formed by a similar formation mechanism.
In addition, the accretion model is also used to explain the formation of the outer rings in Hoag-type galaxies 
\citep{Schweizer1987,FinkelmanBrosch2011}. In Hoag's Object, the complete regularity of the central body is suggested 
to be a result of a recent accretion event \citep{Schweizer1987}. Although PGC 1000714 is not a true Hoag-type galaxy 
due to the inner diffuse ring, the outer ring may be formed by similar mechanism, based on the characteristic similarities.
Table 4 summarizes our comparison between PGC 1000714, the narrow PRG AM 2020-504 \citep{Iodice2002}, the double ringed PRG ESO 474-G26 
\citep{Spavone2012}, Hoag's Object \citep{FinkelmanBrosch2011,Finkelman2011}, and UGC 4599 (a Hoag-type galaxy 
defined by \citet{FinkelmanBrosch2011}).

Given that the colours of the outer ring imply a 0.13 Gyr stellar population, regularity of the central body of PGC 1000714 with 
no tidal signature implies that the galaxy could have undergone a recent accretion event as proposed in PRGs and Hoag-type galaxies. It is possible the gas 
could have been accreted from a single event such as a passing gas-rich dwarf companion. Previous studies of gas-rich early-type galaxies have 
shown that nearby companions are often missed by optical catalogues or show no sign of an optical counterpart \citep[e.g.,][]{Serra2006,Struve2010}. 
A study on the HI environment of PGC 1000714 is therefore needed to complement our study.  

\begin{table*}
\begin{minipage}{150mm}
\centering
\caption{Comparison between PGC 1000714, narrow PRG AM 2020-504, the double ringed-PRG ESO 474-G26, Hoag's Object, and UGC 4599.}
\label{T4}
\begin{tabular}{l|c|c|c|c|c} 
          \hline
           {}  & PGC 1000714 & AM 2020-504 \footnote{\citet{Iodice2002,Iodice2014}} & ESO 474-G26 \footnote{\citet{Spavone2012,Reshetnikov2005}} & Hoag's Object \footnote{\citet{FinkelmanBrosch2011,Finkelman2011}} &  UGC 4599 \footnote{\citet{FinkelmanBrosch2011}} \\
          \hline
           CORE\\
          \hline
           R$_{e}$ (kpc)         & 2.4             & 4.0 \footnote{\citet{Arna1993}}   & 15.3                & 2.0               & 3.2   \\
           $b/a$                & $\sim0.85$       &$\sim0.75$                        &$\sim1$           & $\sim1$            & $\sim0.9$     \\
           m$_{B}$               & $16.36\pm0.06$  & 16.63                            & $13.94\pm0.06$       &...                 &...           \\
           m$_{V}$               & $15.39\pm0.11$   & ...                              &...                   & $16.42\pm0.02$     & $14.26\pm0.02$  \\
           M$_{V}$               & $-19.43\pm0.02$  & ...                              &...                   & $-19.8\pm0.02$     & $-17.9\pm0.02$  \\
           m$_{J}$               & $13.34\pm0.52$  & 13.50                            & $12.51\pm0.02$       & $13.58\pm0.05$     & $11.98\pm0.04$ \\
           V-R                  & $0.2\pm0.04$    &...                               &...                   & $0.86\pm0.04$       & $0.68\pm0.04$   \\
           R-I                  & $1.1\pm0.05$    &...                               &...                   & $0.42\pm0.04$       & $0.38\pm0.04$ \\
           B-K                  & $4.2\pm0.05$    & 3.98                             & $2.55\pm0.08$        & ...                 & ...            \\
           J-K                  & $0.95\pm0.05$   & 1.04                             & $1.12\pm0.04$        & $0.96\pm0.14$       & $0.81\pm0.14$ \\
           Age (Gyr)            & $5.5\pm1.0$     & 3-5                              & 1                    &   $ > 10 $          &     $> 5$  \\
          \hline
           RING 1 \\
          \hline 
           Radius [R$_{1}$] (kpc)   & 15.6                      & $\sim3$              & $\sim29$               &  15.5            &   7.7  \\
           R$_{1}$/R$_{e}$           & 6.5                      & 0.75                 & $\sim1.9$              & 7.8              &  2.4   \\
           m$_{B}$                  & $18.13\pm0.09$           &$17.96-18.40$           & $15.50-16.97$         &...              & ...     \\
           m$_{V}$                  & $17.72\pm0.27$           &...                     &...                   & $16.12\pm0.07$    & $15.18\pm0.02$      \\
           M$_{V}$                  & $-17.07\pm0.02$          &...                     &...                   & $-20.1\pm0.02$    & $-17\pm0.02$        \\
           NUV-V                   & $2.42\pm0.21$            &...                     &...                   & $2.44\pm0.07$     & $1.97\pm0.05$       \\
           V-R                     & $0.21\pm0.04$            &...                     &...                   & $0.51\pm0.10$     & $0.16\pm0.04$        \\
           R-I                     & $0.59\pm0.05$            &...                     &...                   & $0.26\pm0.12$     & $0.34\pm0.04$        \\
           Age (Gyr)               & 0.13                     & 1                      & 0.1-0.03              & $< 2$             &   $\sim1$    \\
           \hline
           RING 2 \\
           \hline
           Radius [R$_{2}$] (kpc)   & $\sim6$                     & ...                  &  $\sim19$              &...                      &...\\
           R$_{2}$/R$_{e}$           & $\sim2.5$                    & ...                  & $\sim1.2$             &...                      &...  \\ 
           m$_{B}$                  & $20.85\pm0.33$               & ...                 & $15.30 - 16.67$         &...                     &...\\
         \hline
        \end{tabular}
\end{minipage}
\end{table*} 

\section{Conclusions}

We present here a photometric study of PGC 1000714, a galaxy with a fairly superficial resemblance to Hoag's Object. The detailed 
2D image analysis reveals that PGC 1000714 is a genuine non-barred galaxy in a low-density environment with two rings. Even with 
the \textit{NIR} bands, we have failed to detect any trace of a bar-like structure. The geometric properties derived from
the optical images, in particular $\epsilon$ and PA, stay constant along the central core. The smooth profiles at the outer 
region imply that the system has settled to its current configuration. The nearly round central body follows closely a r$^{1/4}$ 
light profile for more than 5.0 mag almost all the way to the centre. Based on the 2D $B-I$ colour index map and the photometric 
decomposition presented in this work, we classified the galaxy as a double ringed elliptical (E2), which was misclassified as (R)SAa in NED. 
Without the kinematic data, we cannot confirm or reject the possibility of PGC 1000714 being a PRG. If the galaxy is really a PRG,
it is a very peculiar example of this unique class of objects. If it is not a PRG, this current work has revealed for the first time that 
an elliptical galaxy has two fairly round rings. Studying such peculiar galaxies is important to address how different kinds of interactions
(i.e. internal or external such as galaxy-galaxy and galaxy-environment) lend to different galaxy morphologies. 
 
A number of formation scenarios are discussed based on the photometric results. A careful inspection of the colour profiles 
reveals a fairly constant colour index along the core. The inner ring is very faint in blue light and gets brighter towards redder 
light while the outer ring is very bright in UV light and gets fainter towards redder light. Our SED fitting suggests the galaxy is fairly young 
such that the core is 5.5 Gyr old while the outer ring is just 0.13 Gyr old. The outer ring is well fitted by a stellar population 
with solar metallicity by an instantaneous burst scenario. However, the best-fitting solution for the central body is obtained 
by an exponentially decaying SFH with solar metallicity. This is consistent with the photometric properties of elliptical galaxies 
in the local Universe. Based on the colour of the rings, we expect different formation histories for the inner and outer ring.
To recover the formation history of the inner ring, we need infrared data of PGC 1000714 with better resolution.
Among several ways to create the outer ring in a non-barred galaxy, we conclude that a recent accretion 
event such as accretion from a gas-rich dwarf galaxy is the most plausible scenario that accounts for the observed properties 
of the galaxy. Spectroscopic data and a study of the HI environment  of PGC 1000714 are needed to say more about the evolution 
 of this galaxy in a low-density environment.

\section*{Acknowledgements}
We thank Noah Brosch for his valuable comments and suggestions which have considerably contributed in improving our paper. 

This research has made use of the NASA/IPAC Extragalactic Database (NED) which is operated by the Jet Propulsion Laboratory, 
California Institute of Technology, under contract with National Aeronautics and Space Administration. The Mikulski Archive 
for Space Telescopes (MAST) public access site is used for retrieving Galaxy Evolution Explorer (GALEX) Release 7 data products. 
STScl is operated by the Association of Universities for Research in Astronmy, Inc., under NASA contract NAS5-26555. Support 
for MAST for non-HST data is provided by the NASA Office of Space Science via grant NNX09AF08G and by other grants and contacts.
This publication also makes use of the data products 
from the Two Micron All Sky Survey, which is a joint project of the University of Massachusetts and the Infrared Processing 
and Analysis Center/California Institute of Technology, funded by the National Aeronautics and Space Administration and the 
National Science Foundation. This publication also makes use of data products from the Wide-field Infrared Survey Explorer, 
which is a joint project of the University of California, Los Angeles, and the Jet Propulsion Laboratory/California Institute 
of Technology, funded by the National Aeronautics and Space Administration.

%%%%%%%%%%%%%%%%%%%%%%%%%%%%%%%%%%%%%%%%%%%%%%%%%%

%%%%%%%%%%%%%%%%%%%% REFERENCES %%%%%%%%%%%%%%%%%%

% The best way to enter references is to use BibTeX:

\bibliographystyle{mn2e}
\bibliography{reference} % if your bibtex file is called reference.bib

%%%%%%%%%%%%%%%%%%%%%%%%%%%%%%%%%%%%%%%%%%%%%%%%%%

% Don't change these lines
\bsp	% typesetting comment
\label{lastpage}
\end{document}